%Paper: hep-th/9308105
%From: Jose Miguel Figueroa-O'Farrill <J.M.Figueroa-O'Farrill@qmw.ac.uk>
%Date: Sun, 22 Aug 93 18:41:04 +0100

%%%%%%%%%%%%%%%%%%%%%%%%%%%%%%%%%%%%%%%%%%%%%%%%%%%%%%%%%%%%%%%%
%
%   This is the Plain TeX file for
%
%
%   Constructing N=2 Superconformal Algebras
%
%       out of N=1 Affine Lie Algebras
%
%    by
%
%       J.M. Figueroa-O'Farrill.
%
%  If you encounter any problems compiling this file, please
%
%  send me some e-mail.
%
%
%%%%%%%%%%%%%%%%%%%%%%%%%%%%%%%%%%%%%%%%%%%%%%%%%%%%%%%%%%%%%
%%%These are the macros for submission of papers to hep-th%%%
%%%The default setting is 12pt and 1 page/side but in the%%%%
%%%future it may allow people to choose also 10 pt and%%%%%%%
%%%2 pages/side.%%%%%%%%%%%%%%%%%%%%%%%%%%%%%%%%%%%%%%%%%%%%%
%%%%%%%%%%%%%%%%%%%%%%%%%%%%%%%%%%%%%%%%%%%%%%%%%%%%%%%%%%%%%
%
\def\unlockat{\catcode`\@=11}
\def\lockat{\catcode`\@=12}
\unlockat
\def\d@f@ult{} \newif\ifamsfonts \newif\ifafour
%
% \def\m@ssage{\immediate\write16}  \m@ssage{}
% \m@ssage{hep-th preprint macros.  Last modified 16/10/92 (jmf).}
% \message{These macros work with AMS Fonts 2.1 (available via ftp from}
% \message{e-math.ams.com).  If you have them simply hit "return"; if}
% \message{you don't, type "n" now: }
% \endlinechar=-1  %don't add spaces at end of line
% \read-1 to\@nswer
% \endlinechar=13
% \ifx\@nswer\d@f@ult\amsfontstrue
%     \m@ssage{(Will load AMS fonts.)}
% \else\amsfontsfalse\m@ssage{(Won't load AMS fonts.)}\fi
% %
% \message{The default papersize is A4.  If you use US 8.5" x 11"}
% \message{type an "a" now, else just hit "return": }
% \endlinechar=-1  %don't add spaces at end of line
% \read-1 to\@nswer
% \endlinechar=13
% \ifx\@nswer\d@f@ult\afourtrue
%     \m@ssage{(Using A4 paper.)}
% \else\afourfalse\m@ssage{(Using US 8.5" x 11".)}\fi
% %
%\nonstopmode
%
%%%%%%%%%%%%%%%%%%%%%%
%%%Font definitions%%%
%%%%%%%%%%%%%%%%%%%%%%
%

\font\twelverm=cmr12
\font\ninerm=cmr9
\font\sixrm=cmr6
\font\fourteenbf=cmbx12 scaled\magstep1
\font\twelvebf=cmbx12
\font\ninebf=cmbx9
\font\sixbf=cmbx6
\font\fourteeni=cmmi12 scaled\magstep1      \skewchar\fourteeni='177
\font\twelvei=cmmi12                        \skewchar\twelvei='177
\font\ninei=cmmi9                           \skewchar\ninei='177
\font\sixi=cmmi6                            \skewchar\sixi='177
\font\fourteensy=cmsy10 scaled\magstep2     \skewchar\fourteensy='60
\font\twelvesy=cmsy10 scaled\magstep1       \skewchar\twelvesy='60
\font\ninesy=cmsy9                          \skewchar\ninesy='60
\font\sixsy=cmsy6                           \skewchar\sixsy='60
\font\fourteenex=cmex10 scaled\magstep2
\font\twelveex=cmex10 scaled\magstep1

\ifamsfonts
   \font\ninex=cmex9
   
   \font\sixex=cmex7 at 6pt
   
\else
   \font\ninex=cmex10 at 9pt
   
   \font\sixex=cmex10 at 6pt
   
\fi
\font\fourteensl=cmsl10 scaled\magstep2
\font\twelvesl=cmsl10 scaled\magstep1

\font\sevensl=cmsl10 at 7pt
\font\sixsl=cmsl10 at 6pt

\font\fourteenit=cmti12 scaled\magstep1
\font\twelveit=cmti12

\font\fourteentt=cmtt12 scaled\magstep1
\font\twelvett=cmtt12
\font\fourteencp=cmcsc10 scaled\magstep2
\font\twelvecp=cmcsc10 scaled\magstep1

\ifamsfonts
   
\else
   
\fi
\newfam\cpfam
\font\fourteenss=cmss12 scaled\magstep1
\font\twelvess=cmss12
\font\tenss=cmss10
\font\niness=cmss9

\font\sevenss=cmss8 at 7pt
\font\sixss=cmss8 at 6pt
\newfam\ssfam
\newfam\msafam \newfam\msbfam \newfam\eufam
\ifamsfonts
 \font\fourteenmsa=msam10 scaled\magstep2
 \font\twelvemsa=msam10 scaled\magstep1
 \font\tenmsa=msam10
 \font\ninemsa=msam9
 \font\sevenmsa=msam7
 \font\sixmsa=msam6
 \font\fourteenmsb=msbm10 scaled\magstep2
 \font\twelvemsb=msbm10 scaled\magstep1
 \font\tenmsb=msbm10
 \font\ninemsb=msbm9
 \font\sevenmsb=msbm7
 \font\sixmsb=msbm6
 \font\fourteeneu=eufm10 scaled\magstep2
 \font\twelveeu=eufm10 scaled\magstep1
 \font\teneu=eufm10
 \font\nineeu=eufm9
 
 \font\seveneu=eufm7
 \font\sixeu=eufm6
 \def\hexnumber@#1{\ifnum#1<10 \number#1\else
  \ifnum#1=10 A\else\ifnum#1=11 B\else\ifnum#1=12 C\else
  \ifnum#1=13 D\else\ifnum#1=14 E\else\ifnum#1=15 F\fi\fi\fi\fi\fi\fi\fi}
 \def\hexmsa{\hexnumber@\msafam}
 \def\hexmsb{\hexnumber@\msbfam} 
\fi
\newdimen\b@gheight             \b@gheight=12pt
\newcount\f@ntkey               \f@ntkey=0
\def\f@m{\afterassignment\samef@nt\f@ntkey=}
\def\samef@nt{\fam=\f@ntkey \the\textfont\f@ntkey\relax}
\def\rm{\f@m0 }
\def\mit{\f@m1 }
\def\cal{\f@m2 }
\def\it{\f@m\itfam}
\def\sl{\f@m\slfam}
\def\bf{\f@m\bffam}
\def\tt{\f@m\ttfam}
\def\caps{\f@m\cpfam}
\def\ssf{\f@m\ssfam}
\ifamsfonts
 \def\msa{\f@m\msafam}
 \def\msb{\f@m\msbfam} \let\bb=\msb
 \def\eu{\f@m\eufam}
\else
 \let \bb=\bf \let\eu=\bf
\fi
\def\fourteenpoint{\relax
    \textfont0=\fourteencp          \scriptfont0=\tenrm
      \scriptscriptfont0=\sevenrm
    \textfont1=\fourteeni           \scriptfont1=\teni
      \scriptscriptfont1=\seveni
    \textfont2=\fourteensy          \scriptfont2=\tensy
      \scriptscriptfont2=\sevensy
    \textfont3=\fourteenex          \scriptfont3=\twelveex
      \scriptscriptfont3=\tenex
    \textfont\itfam=\fourteenit     \scriptfont\itfam=\tenit
    \textfont\slfam=\fourteensl     \scriptfont\slfam=\tensl
      \scriptscriptfont\slfam=\sevensl
    \textfont\bffam=\fourteenbf     \scriptfont\bffam=\tenbf
      \scriptscriptfont\bffam=\sevenbf
    \textfont\ttfam=\fourteentt
    \textfont\cpfam=\fourteencp
    \textfont\ssfam=\fourteenss     \scriptfont\ssfam=\tenss
      \scriptscriptfont\ssfam=\sevenss
    \ifamsfonts
       \textfont\msafam=\fourteenmsa     \scriptfont\msafam=\tenmsa
         \scriptscriptfont\msafam=\sevenmsa
       \textfont\msbfam=\fourteenmsb     \scriptfont\msbfam=\tenmsb
         \scriptscriptfont\msbfam=\sevenmsb
       \textfont\eufam=\fourteeneu     \scriptfont\eufam=\teneu
         \scriptscriptfont\eufam=\seveneu \fi
    \samef@nt
    \b@gheight=14pt
    \setbox\strutbox=\hbox{\vrule height 0.85\b@gheight
                                depth 0.35\b@gheight width\z@ }}
\def\twelvepoint{\relax
    \textfont0=\twelverm          \scriptfont0=\ninerm
      \scriptscriptfont0=\sixrm
    \textfont1=\twelvei           \scriptfont1=\ninei
      \scriptscriptfont1=\sixi
    \textfont2=\twelvesy           \scriptfont2=\ninesy
      \scriptscriptfont2=\sixsy
    \textfont3=\twelveex          \scriptfont3=\ninex
      \scriptscriptfont3=\sixex
    \textfont\itfam=\twelveit    %\scriptfont\itfam=\nineit
    \textfont\slfam=\twelvesl    %\scriptfont\slfam=\ninesl
      \scriptscriptfont\slfam=\sixsl
    \textfont\bffam=\twelvebf     \scriptfont\bffam=\ninebf
      \scriptscriptfont\bffam=\sixbf
    \textfont\ttfam=\twelvett
    \textfont\cpfam=\twelvecp
    \textfont\ssfam=\twelvess     \scriptfont\ssfam=\niness
      \scriptscriptfont\ssfam=\sixss
    \ifamsfonts
       \textfont\msafam=\twelvemsa     \scriptfont\msafam=\ninemsa
         \scriptscriptfont\msafam=\sixmsa
       \textfont\msbfam=\twelvemsb     \scriptfont\msbfam=\ninemsb
         \scriptscriptfont\msbfam=\sixmsb
       \textfont\eufam=\twelveeu     \scriptfont\eufam=\nineeu
         \scriptscriptfont\eufam=\sixeu \fi
    \samef@nt
    \b@gheight=12pt
    \setbox\strutbox=\hbox{\vrule height 0.85\b@gheight
                                depth 0.35\b@gheight width\z@ }}
\twelvepoint
%
%%%%%%%%%%%%%%%%%
%%%Basic skips%%%
%%%%%%%%%%%%%%%%%
%
\baselineskip = 15pt plus 0.2pt minus 0.1pt %was 20pt ...
\lineskip = 1.5pt plus 0.1pt minus 0.1pt
\lineskiplimit = 1.5pt
\parskip = 6pt plus 2pt minus 1pt
\interlinepenalty=50
\interfootnotelinepenalty=5000
\predisplaypenalty=9000
\postdisplaypenalty=500
\hfuzz=1pt
\vfuzz=0.2pt
\dimen\footins=24 truecm % 8 truein in SB
\ifafour
 \hsize=16cm \vsize=22cm
\else
 \hsize=6.5in \vsize=9in
\fi
%
%%%%%%%%%%%%%%%
%%%Footnotes%%%
%%%%%%%%%%%%%%%
%
\skip\footins=\medskipamount
\newcount\fnotenumber
\def\clearfnotenumber{\fnotenumber=0} \clearfnotenumber
\def\fnote{\global\advance\fnotenumber by1 \generatefootsymbol
 \footnote{$^{\footsymbol}$}}
\def\fd@f#1 {\xdef\footsymbol{\mathchar"#1 }}
\def\generatefootsymbol{\iffrontpage\ifcase\fnotenumber
\or \fd@f 279 \or \fd@f 27A \or \fd@f 278 \or \fd@f 27B
\else  \fd@f 13F \fi
\else\xdef\footsymbol{\the\fnotenumber}\fi}
%
%%%%%%%%%%%%%%%%%%%%%%%%%%%%%
%%%Sections and Appendices%%%
%%%%%%%%%%%%%%%%%%%%%%%%%%%%%
%
\newcount\secnumber \newcount\appnumber
\def\clearappnumber{\appnumber=64} \def\clearsecnumber{\secnumber=0}
\clearsecnumber \clearappnumber
\newif\ifs@c % this is true if within a section as opposed to an appendix
\newif\ifs@cd % this is true if the article is being section'd
\s@cdtrue % this is the default
\def\unsectioned{\s@cdfalse\let\section=\subsection}
\newskip\sectionskip         \sectionskip=\medskipamount
\newskip\headskip            \headskip=8pt plus 3pt minus 3pt
\newdimen\sectionminspace    \sectionminspace=10pc
\def\Titlestyle#1{\par\begingroup \interlinepenalty=9999
     \leftskip=0.02\hsize plus 0.23\hsize minus 0.02\hsize
     \rightskip=\leftskip \parfillskip=0pt
     \advance\baselineskip by 0.5\baselineskip%this is a test...
     \hyphenpenalty=9000 \exhyphenpenalty=9000
     \tolerance=9999 \pretolerance=9000
     \spaceskip=0.333em \xspaceskip=0.5em
     \fourteenpoint
  \noindent #1\par\endgroup }
\def\titlestyle#1{\par\begingroup \interlinepenalty=9999
     \leftskip=0.02\hsize plus 0.23\hsize minus 0.02\hsize
     \rightskip=\leftskip \parfillskip=0pt
     \hyphenpenalty=9000 \exhyphenpenalty=9000
     \tolerance=9999 \pretolerance=9000
     \spaceskip=0.333em \xspaceskip=0.5em
     \fourteenpoint
   \noindent #1\par\endgroup }
\def\spacecheck#1{\dimen@=\pagegoal\advance\dimen@ by -\pagetotal
   \ifdim\dimen@<#1 \ifdim\dimen@>0pt \vfil\break \fi\fi}
\def\section#1{\cleareqnumber \s@ctrue \global\advance\secnumber by1
   \par \ifnum\the\lastpenalty=30000\else
   \penalty-200\vskip\sectionskip \spacecheck\sectionminspace\fi
   \noindent {\caps\enspace\S\the\secnumber\quad #1}\par
   \nobreak\vskip\headskip \penalty 30000 }
\def\undertext#1{\vtop{\hbox{#1}\kern 1pt \hrule}}
\def\subsection#1{\par
   \ifnum\the\lastpenalty=30000\else \penalty-100\smallskip
   \spacecheck\sectionminspace\fi
   \noindent\undertext{#1}\enspace \vadjust{\penalty5000}}

\def\appendix#1{\cleareqnumber \s@cfalse \global\advance\appnumber by1
   \par \ifnum\the\lastpenalty=30000\else
   \penalty-200\vskip\sectionskip \spacecheck\sectionminspace\fi
   \noindent {\caps\enspace Appendix \char\the\appnumber\quad #1}\par
   \nobreak\vskip\headskip \penalty 30000 }
\def\ack{\par\penalty-100\medskip \spacecheck\sectionminspace
   \line{\fourteencp\hfil ACKNOWLEDGEMENTS\hfil}%
\nobreak\vskip\headskip }
\def\refs{\begingroup \par\penalty-100\medskip \spacecheck\sectionminspace
   \line{\fourteencp\hfil REFERENCES\hfil}%
\nobreak\vskip\headskip \frenchspacing }
\def\endrefs{\par\endgroup}
%--- Note added
%
%%%%%%%%%%%%%%%%%%%%%%%%%%%%%%%%%
%%%Running heads and footlines%%%
%%%%%%%%%%%%%%%%%%%%%%%%%%%%%%%%%
%
\newif\iffrontpage \frontpagefalse
\headline={\hfil}
\footline={\iffrontpage\hfil\else \hss\twelverm
-- \folio\ --\hss \fi }
%
%%%%%%%%%%%%%%%%
%%%Title page%%%
%%%%%%%%%%%%%%%%
%
\newskip\frontpageskip \frontpageskip=12pt plus .5fil minus 2pt
\def\titlepage{\global\frontpagetrue\hrule height\z@ \relax
               \pubblock\relax }
\def\endtitlepage{\vfil\break\clearfnotenumber\frontpagefalse}
\def\title#1{\vskip\frontpageskip\Titlestyle{\caps #1}\vskip3\headskip}
\def\author#1{\vskip.5\frontpageskip\titlestyle{\caps #1}\nobreak}
\def\and{\par\kern 5pt \centerline{\sl and}}

\def\address#1{\par\kern 5pt\titlestyle{\it #1}}
\def\andaddress{\par\kern 5pt \centerline{\sl and} \address}

\def\abstract#1{\par\dimen@=\prevdepth \hrule height\z@ \prevdepth=\dimen@
   \vskip\frontpageskip\spacecheck\sectionminspace
   \centerline{\fourteencp ABSTRACT}\vskip\headskip
   {\noindent #1}}

\def\email#1{\fnote{\tentt e-mail: #1\hfill}}
\def\newaddress#1{\fnote{\tenrm #1\hfill}}
%
%%%%%%%%%%%%%%%%%%%%
%%%some addresses%%%
%%%%%%%%%%%%%%%%%%%%
%

%
\def\Bonn{\address{%
   Physikalisches Institut der Universit\"at Bonn\break
  Nu{\ss}allee 12, D--53115 Bonn, GERMANY}}
%

%

%
%%%%%%%%%%%%%%%%
%%%References%%%
%%%%%%%%%%%%%%%%
%
\newcount\refnumber \def\clearrefnumber{\refnumber=0}  \clearrefnumber
\newwrite\R@fs                              %This opens a file .refs with
\immediate\openout\R@fs=\jobname.refs %the references in order of
                                            %appearance.
\def\closerefs{\immediate\closeout\R@fs} %close file so that TeX can read it
\def\refsout{\closerefs\refs
\unlockat
\input\jobname.refs
\lockat
\endrefs}
\def\refitem#1{\item{{\bf #1}}}%just bolds it so that \bf does not expand
\def\ifundefined#1{\expandafter\ifx\csname#1\endcsname\relax}
\def\[#1]{\ifundefined{#1R@FNO}%
\global\advance\refnumber by1%
\expandafter\xdef\csname#1R@FNO\endcsname{[\the\refnumber]}%
\immediate\write\R@fs{\noexpand\refitem{\csname#1R@FNO\endcsname}%
\noexpand\csname#1R@F\endcsname}\fi{\bf \csname#1R@FNO\endcsname}}
\def\refdef[#1]#2{\expandafter\gdef\csname#1R@F\endcsname{{#2}}}
%
%%%%%%%%%%%%%%%
%%%Equations%%%
%%%%%%%%%%%%%%%
%
\newcount\eqnumber \def\cleareqnumber{\eqnumber=0}
\newif\ifal@gn \al@gnfalse  % this is true if within an \eqalignno
\def\veqnalign#1{\al@gntrue \vbox{\eqalignno{#1}} \al@gnfalse}
\def\eqnalign#1{\al@gntrue \eqalignno{#1} \al@gnfalse}
\def\(#1){\relax%
\ifundefined{#1@Q}
 \global\advance\eqnumber by1
 \ifs@cd
  \ifs@c
   \expandafter\xdef\csname#1@Q\endcsname{{%
\noexpand\rm(\the\secnumber .\the\eqnumber)}}
  \else
   \expandafter\xdef\csname#1@Q\endcsname{{%
\noexpand\rm(\char\the\appnumber .\the\eqnumber)}}
  \fi
 \else
  \expandafter\xdef\csname#1@Q\endcsname{{\noexpand\rm(\the\eqnumber)}}
 \fi
 \ifal@gn
    & \csname#1@Q\endcsname
 \else
    \eqno \csname#1@Q\endcsname
 \fi
\else%
\csname#1@Q\endcsname\fi\global\let\@Q=\relax}
%
%%%%%%%%%%%%%%%%%
%%%Mathematica%%%
%%%%%%%%%%%%%%%%%
%
\newif\ifm@thstyle \m@thstylefalse
\def\mathstyle{\m@thstyletrue}
\def\proclaim#1#2\par{\smallbreak\begingroup%        small --> med???
\advance\baselineskip by -0.25\baselineskip%
\advance\belowdisplayskip by -0.35\belowdisplayskip%
\advance\abovedisplayskip by -0.35\abovedisplayskip%
    \noindent{\caps#1.\enspace}{#2}\par\endgroup%
\smallbreak}%--- defs, thms, ...                     small --> med???
\def\m@kem@th<#1>#2#3{%
\ifm@thstyle \global\advance\eqnumber by1
 \ifs@cd
  \ifs@c
   \expandafter\xdef\csname#1\endcsname{{%
\noexpand #2\ \the\secnumber .\the\eqnumber}}
  \else
   \expandafter\xdef\csname#1\endcsname{{%
\noexpand #2\ \char\the\appnumber .\the\eqnumber}}
  \fi
 \else
  \expandafter\xdef\csname#1\endcsname{{\noexpand #2\ \the\eqnumber}}
 \fi
 \proclaim{\csname#1\endcsname}{#3}
\else
 \proclaim{#2}{#3}
\fi}
\def\Thm<#1>#2{\m@kem@th<#1M@TH>{Theorem}{\sl#2}}%--- Theorem
\def\Prop<#1>#2{\m@kem@th<#1M@TH>{Proposition}{\sl#2}}%--- Proposition
\def\Def<#1>#2{\m@kem@th<#1M@TH>{Definition}{\rm#2}}%--- Definition
\def\Lem<#1>#2{\m@kem@th<#1M@TH>{Lemma}{\sl#2}}%--- Lemma
\def\Cor<#1>#2{\m@kem@th<#1M@TH>{Corollary}{\sl#2}}%--- Corollary
\def\Conj<#1>#2{\m@kem@th<#1M@TH>{Conjecture}{\sl#2}}%--- Conjecture
\def\Rmk<#1>#2{\m@kem@th<#1M@TH>{Remark}{\rm#2}}%--- Remark
\def\Exm<#1>#2{\m@kem@th<#1M@TH>{Example}{\rm#2}}%--- Example
\def\Qry<#1>#2{\m@kem@th<#1M@TH>{Query}{\it#2}}%--- Query
%
%--- Proof
%

%
\def\<#1>{\csname#1M@TH\endcsname}
%
%%%%%%%%%%%%%%%%%%%
%%%Abbreviations%%%
%%%%%%%%%%%%%%%%%%%
%
\def\ref#1{{\bf [#1]}}%--- [ref]
%--- et al.
%--- i.e.
%--- e.g.
%--- Cf.
%--- cf.
 %--- double left quote
%--- th as in fifth
\def\nl{\hfil\break}%--- new line
%
%%%%%%%%%%%%%%%%%
%%%Mathematics%%%
%%%%%%%%%%%%%%%%%
%
%--- def over =
%--- Halmos Q.E.D.

%--- implies
%--- is implied by
%--- if and only if
\def\lapprox{\hbox{\lower3pt\hbox{$\buildrel<\over\sim$}}}% approx lt
\def\gapprox{\hbox{\lower3pt\hbox{$\buildrel<\over\sim$}}}% approx gt
\def\quotient#1#2{#1/\lower0pt\hbox{${#2}$}}%--- factor objects
\def\fr#1/#2{\mathord{\hbox{${#1}\over{#2}$}}}
\ifamsfonts
 \mathchardef\empty="0\hexmsb3F %--- better empty set than \emptyset
 \mathchardef\lsemidir="2\hexmsb6E % semidirect |x
 \mathchardef\rsemidir="2\hexmsb6F % semidirect x|
\else
 \let\empty=\emptyset
 \def\lsemidir{\mathbin{\hbox{\hskip2pt\vrule height 5.7pt depth -.3pt
    width .25pt\hskip-2pt$\times$}}}
 \def\rsemidir{\mathbin{\hbox{$\times$\hskip-2pt\vrule height 5.7pt
    depth -.3pt width .25pt\hskip2pt}}}
\fi
%
%--- injective map
%--- surjective map
%--- bijective map
%--- mapping
%--- long mapping
%--- isom over -->
%--- just an abbrev.
%

%
 %--- commutative diagram macro
 %--- map in complex
%
 %--- reals
\def\comps{\mathord{\bb C}} %--- complex nos.
 %--- quaternions
 %--- integers
 %--- rationals
 %--- naturals
 %--- ground field
%
%--- Hom(omorphisms)
%--- tr(ace)
%--- Tr(ace)
%--- End(omorphisms)
%--- Mor(phisms)
%--- Aut(omorphisms)
%--- aut(omorphisms)
%--- supertrace
%--- superdeterminant
%--- kernel
%--- cokernel
%--- image
%
\def\underrightarrow#1{\vtop{\ialign{##\crcr
      $\hfil\displaystyle{#1}\hfil$\crcr
      \noalign{\kern-\p@\nointerlineskip}
      \rightarrowfill\crcr}}} %--- modification of \overrightarrow
\def\underleftarrow#1{\vtop{\ialign{##\crcr
      $\hfil\displaystyle{#1}\hfil$\crcr
      \noalign{\kern-\p@\nointerlineskip}
      \leftarrowfill\crcr}}}  %--- modification of \overleftarrow

\def\comm#1#2{\left[#1\, ,\,#2\right]}%--- [ , ]
%--- { , }
%--- [ , }
%
%--- Lie derivative
%--- vartnl derivative
%--- partial derivative
%--- full derivative
%
%%%%%%%%%%%%%%
%%%Journals%%%
%%%%%%%%%%%%%%
%

\def\NPB#1#2#3{{\sl Nucl. Phys.} {\bf B#1} (#2) #3}

\def\CMP#1#2#3{{\sl Comm. Math. Phys.} {\bf #1} (#2) #3}

\def\PLB#1#2#3{{\sl Phys. Lett.} {\bf #1B} (#2) #3}
\def\JMP#1#2#3{{\sl J. Math. Phys.} {\bf #1} (#2) #3}

\def\IJMPA#1#2#3{{\sl Int. J. Mod. Phys.} {\bf A#1} (#2) #3}
\def\IJMPC#1#2#3{{\sl Int. J. Mod. Phys.} {\bf C#1} (#2) #3}

\def\MPLA#1#2#3{{\sl Mod. Phys. Lett.} {\bf A#1} (#2) #3}
\def\JETP#1#2#3{{\sl Sov. Phys. JETP} {\bf #1} (#2) #3}

\lockat

%%%%%%%%%%%%%%%%%%%%%%%%%%%%%%%%%%%%%%%%%%%%%%%%%%%%%%%%%%%%%
%
%   These are the local macros for BONN-HE-93-Gen N=2
%
\def\W{\mathord{\ssf W}}
\def\WB{\mathord{\ssf WB}}
\def\ope[#1][#2]{{{#2}\over{\ifnum#1=1 {z-w} \else {(z-w)^{#1}}\fi}}}

\def\reg{\mathord{\hbox{reg.}}}
\def\ad{\mathord{\hbox{ad}}}

\def\gg{{\mathord{\eu g}}}

\def\gghat{\widehat{\gg}}

\let\tensor=\otimes

\def\fr#1/#2{\mathord{\hbox{${#1}\over{#2}$}}}

\let\d=\partial
\def\hepth/#1/{{\tt hep-th/#1}}
\def\using[#1]{&\hbox{by }#1}

\def\comm[#1,#2]{\mathord{\left[#1\mathbin{,}#2\right]}}
\def\f[#1,#2]{\mathord{{f_{#1}}^{#2}}}
\def\exc[#1,#2]{\mathord{(#1 \leftrightarrow #2)}}
\def\pair<#1,#2>{\mathop{\left\langle #1 \mathbin{,} #2\right\rangle}}
\refdef[getzler]{E. Getzler, {\sl Manin triples and $N{=}2$
superconformal field theory}, \hepth/9307041/.}
\refdef[jmf]{J.M. Figueroa-O'Farrill, {\sl Affine Algebras, $N{=}2$
Superconformal Algebras and Gauged WZNW Models}, \hepth/9306164/, to
appear in {\sl Physics Letters B}.}
\refdef[highlevel]{M.B. Halpern and N.A. Obers, \NPB{345}{1990}{607}.}
\refdef[FMO]{J.M. Figueroa-O'Farrill, N. Mohammedi, and N. Obers, work
in progress.}
\refdef[Kris]{K. Thielemans, \IJMPC{2}{1991}{662}; {\sl Proc. of the
Second Int. Workshop on Software Engineering, Artificial Intelligence
and Expert Systems in High Energy and Nuclear Physics (1992)}, World
Scientific, p. 709.}
\refdef[VME]{M.B. Halpern, {\sl Recent Developments in the Virasoro
Master Equation}, in the proceedings of the Stony Brook conference,
Strings and Symmetries 1991, World Scientific, 1991.}
\refdef[Nouri]{N. Mohammedi, \PLB{260}{1991}{317}.}
\refdef[GHKO]{A. Giveon, M.B. Halpern, E. Kiritsis, and N. Obers,
\IJMPA{7}{1992}{947}.}
\refdef[KSBosonic]{Y. Kazama and H. Suzuki, \MPLA{4}{1989}{235}.}
\refdef[Gepner]{D. Gepner and Z. Qiu, \NPB{285}{1987}{423}.}
\refdef[Ahn]{C. Ahn, S. Chung, and S.H. Tye, \NPB{365}{1991}{191}.}
\refdef[FiRa]{J.M. Figueroa-O'Farrill and E. Ramos,
\NPB{368}{1992}{361}.}
\refdef[HalpernReview]{M.B. Halpern, {\sl Recent developments in the
Virasoro master equation}, Proceedings of the Stony Brook conference
{\sl Strings and Symmetries 1991}, World Scientific 1991.}
\refdef[HK]{M.B. Halpern and E. Kiritsis, \MPLA{4}{1989}{1373},
Erratum \MPLA{4}{1989}{1797}.}
\refdef[SME]{M.B. Halpern and N. Obers, \IJMPA{7}{1992}{7263},
\IJMPA{7}{1992}{3065}, \JMP{32}{1991}{3231}.}
\refdef[leuven]{A. Sevrin, Ph. Spindel, W. Troost and A. Van Proeyen,
\NPB{308}{1988}{662}, \NPB{311}{1988/89}{465}.}
\refdef[TFT]{E. Witten, \CMP{117}{1988}{353}, \CMP{118}{1988}{411},
\NPB{340}{1990}{281};\nl
T. Eguchi and S. Yang, \MPLA{4}{1990}{1693}.}
\refdef[GRS]{B. Gato-Rivera and A.M. Semikhatov, \PLB{293}{1992}{72}\nl
(\hepth/9207004/).}
\refdef[KaSu]{Y. Kazama and H. Suzuki, \PLB{216}{1989}{112},
\NPB{321}{1989}{232}.}
\refdef[vNSS]{P. van Nieuwenhuizen, K. Schoutens, and A. Sevrin,
\CMP{124}{1989}{87}.}
\refdef[Horn]{K. Hornfeck, {\sl Explicit construction of the BRST
charge for $\W_4$}, \hepth/9306019/.}
\refdef[Zhu]{C.-J. Zhu, {\sl The BRST quantization of the nonlinear
$\WB_2$ and $\W_4$ algebras}, \hepth/9306026/.}
\refdef[QDS]{B. Feigin and E. Frenkel, \PLB{246}{1990}{75};\nl
J.M. Figueroa-O'Farrill, \NPB{343}{1990}{450};\nl
J. de Boer and T. Tjin, {\sl The relation between quantum
$\W$-algebras and Lie algebras}, \hepth/9302006/;\nl
A. Sevrin and W. Troost, {\sl Extensions of the Virasoro algebra and
Gauged WZW models}, \hepth/9306033/.}
\refdef[MV]{S. Mukhi and C. Vafa, {\sl Two-dimensional black hole as a
topological coset model of $c{=}1$ string theory}, \hepth/9301083/.}
\refdef[BLNW]{M. Bershadsky, W. Lerche, D. Nemeshansky, and N. Warner,
{\sl Extended $N{=}2$ superconformal structure of gravity and
$\W$-gravity coupled to matter}, \hepth/9211040/.}
\refdef[wiiineqii]{C. Pope, private communication;\nl
J.M. Figueroa-O'Farrill and K. Thielemans, unpublished.}
\refdef[Romans]{L. Romans, \NPB{352}{1991}{829}.}
\refdef[HuSp]{C.M. Hull and B. Spence, \PLB{241}{1990}{357}.}
\refdef[Sing]{N.P. Warner, {\sl Lectures on $N{=}2$ supercoformal
theories and singularity theory}, in {\it Superstrings '89\/}, Trieste
Spring School, April 1989, World Scientific 1990.}
\refdef[Mirror]{See for example {\sl Essays on Mirror Symmetry}, ed.
by S.-T. Yau}
\refdef[Ring]{W. Lerche, C. Vafa, N.P. Warner, \NPB{324}{1989}{427}.}
\refdef[Dave]{D. Montano and J. Sonnenschein, \NPB{324}{1989}{348}.}
\refdef[Park]{S. Parkhomenko, \JETP{75}{1992}{1}.}
\refdef[GKO]{P. Goddard, A. Kent, and D. Olive, \PLB{152}{1985}{88};
\CMP{103}{1986}{105}.}
\refdef[Martin]{A. Giveon and M. Ro{\v c}ek, {\sl On the BRST Operator
Structure of the $N{=}2$ String}, \hepth/9302049/.}

\overfullrule=0pt
%\draftmode
%
\def\pubblock{ \line{\hfil\rm BONN--HE--93--21}
               \line{\hfil\tt hep-th/9308105}
               \line{\hfil\rm August 1993}}
\titlepage
\title{Constructing $N{=}2$ Superconformal Algebras out of $N{=}1$
Affine Lie Algebras}
\author{Jos\'e~M.~Figueroa-O'Farrill
\email{jmf@avzw01.physik.uni-bonn.de}
\newaddress{Address after September 1, 1993: Physics Department, Queen
Mary and\nl Westfield College, London, UK.}}
\Bonn
\abstract{We study the problem of constructing $N{=}2$ superconformal
algebras out of an $N{=}1$ affine Lie algebra.  Following a recent
independent observation of Getzler and the author, we derive a
simplified set of $N{=}2$ master equations, which we then proceed to
solve for the case of $sl(2)$.  There is a unique construction for all
noncritical values of the level, which can be identified as the
Kazama--Suzuki coset associated to the hermitian symmetric space
$SO(3)/SO(2)$.  We also identify the construction with a generalized
parafermionic construction or, after bosonization, with a bosonic
construction of the type analyzed by Kazama and Suzuki.  A mild
generalization of this construction can be associated to any embedding
$sl(2)\subset\gg$.}
\endtitlepage

\section{Introduction}

In the course of constructing $N{=}2$ superconformal algebras (SCAs)
from Lie algebraic data, the following characterization of an $N{=}2$
SCA was recently found \[jmf]\[getzler].  Let $G^\pm(z)$ be fields
satisfying the following operator product expansions:
$$
\eqnalign{
G^\pm(z) G^\pm(w) &= \reg~, \(GGope)\cr
\noalign{\hbox{and}}
G^\pm(z) G^\mp(w) &= \ope[3][2c/3] \pm \ope[2][2J(w)] + \ope[1][2T(w)
\pm \d J(w)] + \reg~, \(GpGmope)\cr}
$$
for some fields $J(z)$, $T(z)$ and constant $c$ {\it defined} by the
above OPE.  If, in addition, $J(z)$ and $G^\pm(z)$ satisfy
$$
J(z)G^\pm(w) = \pm \ope[1][G^\pm(w)] + \reg~, \(JGope)
$$
then it was proven independently in \[jmf] and \[getzler] that
$J$, $G^\pm$, and $T$ obey an $N{=}2$ superconformal algebra with
central charge $c$.  The purpose of this paper is to use this result
to analyze the construction of $N{=}2$ SCAs using $N{=}1$ affine Lie
algebras.

The study of $N{=}2$ SCAs is of fundamental importance.  Among the
conformal field theories, those with $N{=}2$ superconformal symmetry
($N{=}2$ SCFTs) are arguably the most interesting both from the point
of view of applications and from purely structural considerations.  It
was realized rather early in modern string theory that $N{=}2$
superconformal invariance is necessary for spacetime supersymmetry,
whence $N{=}2$ SCFTs span the phenomenologically interesting classical
vacua for string theory; for example, string compactifications on a
Calabi-Yau space.

On the more mathematical side, $N{=}2$ SCFTs have a very rich
algebraic structure which makes them omnipresent in string theory and
topological field theory.  It has been realized for some time now that
to any $N{=}2$ SCFT one can associate a cohomology theory which is
intimately linked to the topological conformal field theory (TCFT) one
obtains after twisting \[TFT].  The cohomology of the
TCFT---generically a topological invariant of some moduli space
\[Dave]---inherits in the $N{=}2$ formulation a natural ring
structure.  This so-called chiral ring\[Ring] is a useful invariant of
the theory and accounts for much of the recent popularity of these
theories---lying, for instance, at the center of the mirror symmetry
revolution\[Mirror].  Through the Landau-Ginzburg description of
$N{=}2$ SCFTs, the chiral ring lets one also make contact with
singularity theory\[Sing].

Conversely, there is a growing body of evidence that suggests that to
any reasonable cohomology theory there is associated an $N{=}2$ SCA.
This was shown explicitly for the first time in the literature in
\[GRS] (see also \[MV]) for the $c{=}1$ noncritical string.  In
\[BLNW] this observation was generalized and argued to be a generic
property of string theories: be it the ``humble'' string, the
superstring, or the $\W$-string.  Explicit constructions of the
$N{=}2$ algebra depend on the model under consideration and were given
in \[BLNW] for the bosonic string (critical and noncritical), the NSR
string (critical and noncritical)---where one actually has an $N{=}3$
SCA---and for the noncritical $\W_3$-string, where one finds an
extension of $N{=}2$ by an $N{=}2$ primary of weight $2$ and charge
$0$; and in \[Martin] for the $N{=}2$ string---where one actually has
an $N{=}4$ SCA.  For the critical $\W_3$ string with matter
representation given by the Romans realization of $\W_3$ in terms of a
free boson and an underlying Virasoro algebra \[Romans] one can also
construct such an $N{=}2$ extended algebra \[wiiineqii] whose central
charge is fixed to either of the two values $c=-18$ or $c=-\fr15/2$.
It is expected that this continues to be the case for other
$\W$-string theories built on other $\W$-algebras; but proving it by
the current means requires knowing (at least the existence of) the
BRST charge for the relevant $\W$-algebra, for which no general
construction is known and hence must be done case by case.  The
computational complexity soon becomes forbidding (even to a computer)
and so far the only other algebras whose BRST charge is known are:
$\W_4$ \[Horn] \[Zhu], $\W(2,4)$ \[Zhu] and some special quadratically
nonlinear algebras \[vNSS].  One can certainly envision a few more
algebras to be reached in the near future, but this is far from a
general existence proof.

Since $\W$-algebras are generically defined via the quantum
Drinfel'd-Sokolov reduction \[QDS], it is widely believed that the BRST
charge and the underlying $N{=}2$ structure should also come induced
from the analogous structures in the affine Lie algebra from which one
reduces.  A first step in this direction was given in \[jmf], where
the underlying $N{=}2$ structure for the BRST complex associated to
the affinization of a semisimple Lie algebra was unveiled.  In
contrast with the BRST complexes in string theory, the construction in
\[jmf] is completely natural and does not depend on any details of the
representation of the affine Lie algebra that one is considering.

Independently and at the same time, Getzler\[getzler] found a very
general class of $N{=}2$ constructions associated to Manin triples.
In doing so he makes explicit an observation of \[Park] concerning the
work in \[leuven].  The Getzler constructions interpolate between the
construction in \[jmf] and (a deformation of) the Kazama--Suzuki coset
construction \[KaSu].

Despite all of these constructions now in existence, much fewer
constructions are known for $N{=}2$ SCFTs than for their $N{=}0$ and
$N{=}1$ counterparts.  For example, no natural Sugawara-type
construction exists for $N{=}2$ SCFTs (although see \[HuSp]) and
already the $N{=}2$ extension of the coset construction \[GKO] imposes
restrictions on the geometry of the coset  manifold, requiring it to
be K\"ahler \[KaSu].  This kind of rigidity is typical of $N{=}2$
constructions and we shall encounter it again in this paper.

Although in contrast to the $N{=}0$ and $N{=}1$ cases, there is no
canonical construction which one is generalizing, we can place this
work in the context of generalized $N{=}2$ superconformal
constructions.  Generalized Virasoro constructions, conceived
originally by Halpern and Kiritsis \[HK], have so far had a short but
lively history which is amply documented in \[HalpernReview].  The
$N{=}1$ case was analyzed originally by \[Nouri] and \[GHKO] and then
further in \[SME].  For the $N{=}2$ case much less in known.  A set of
master equations was derived in an appendix of \[GHKO], who verified
that the Kazama--Suzuki constructions satisfied the equations, but did
not look further.  In this paper we will derive a set of master
equations for ``generalized'' $N{=}2$ constructions which comprise a
subset of the equations in \[GHKO] and generate the remaining ones.
Although the equations have sufficient structure to expect that some
results of a general nature can be obtained for all $\gg$, we will
only solve the case of $\gg=sl(2)$.  We find that there is a unique
construction, which after some cosmetics can be identified  with the
Kazama--Suzuki coset construction associatd to $SO(3)/SO(2)$ or with a
generalized parafermionic construction.  We also identify it with a
bosonic construction the type analyzed by Kazama and Suzuki in
\[KSBosonic].  We also remark a similarity with the BRST complex of
the quantum Drinfel'd--Sokolov reduction that suggests a conjecture
described in the conclusions.

The paper is organized as follows.  In Section 2 we derive the $N{=}2$
master equations associated to an $N{=}1$ affine Lie algebra at level
$k$ and prove that there are no nontrivial solutions for $k=0$.  This
means that we are free to decouple the fermions, which we do in
Section 3 where we derive the simplified master equations.  In Section
4 we solve the master equations for the simplest case $\gg=sl(2)$.
The solution benefits from a geometrization of the master equation
that is particular to $sl(2)$, but makes the underlying mechanism very
transparent. A mild deformation of this construction then applies to
an arbitrary simple Lie algebra and we discuss this in Section 5.  In
Section 6 we identify this construction.  It turns out to be
equivalent to the Kazama--Suzuki coset construction associated to
$SO(3)/SO(2)$, to a (generalized) parafermionic construction and to a
bosonic construction of the type analyzed by Kazama and Suzuki.
Finally in section 7 we summarize our results.

\section{The $N{=}2$ Master Equations}

Let $\gg$ be a simple complex Lie algebra with a fixed invariant
metric $\pair<,>$.  Let us fix a basis $\{X_i\}$ for $\gg$.  Relative
to this basis, $\pair<X_i,X_j> = g_{ij}$ and $\comm[X_i,X_j]=\f[ij,k]
X_k$.  The $N{=}1$ affine Lie algebra (sometimes called a super
Ka{\v c}--Moody algebra, or a Ka{\v c}--Todorov algebra, with some
abuse of notation) associated to $\gg$ is the natural $N{=}1$
supersymmetric extension of the affine Lie algebra $\gghat$.  It is
generated by currents $I_i(z)$ and $\psi_i(z)$ obeying the following
OPEs:
$$
\eqnalign{
I_i(z) I_j(w) &= \ope[2][k g_{ij}] + \ope[1][{f_{ij}}^k I_k(w)] +
\reg~, \(affineKM)\cr
I_i(z) \psi_j(w) &= \ope[1][{f_{ij}}^k \psi_k(w)] + \reg~, \(Jpsi)\cr
\noalign{\hbox{and}}
\psi_i(z) \psi_j(w) &= \ope[1][k g_{ij}] + \reg~, \(fermions)\cr}
$$

It is well-known that for $k\neq 0$ one can decouple the fermions by
introducing currents
$$J_i(z) \equiv I_i(z) - {1\over 2k} {f_{ij}}^k g^{jl}
(\psi_k\psi_l)(z)\(shiftedj)$$
which have a regular OPE with the fermions $\psi_i(z)$.  The
$J_i(z)$ satisfy \(affineKM) with a shifted level $k - \fr1/2
c_{\gg}$, with $c_\gg$ being defined as the eigenvalue of the
quadratic casimir $g^{ij}\ad X_i \ad X_j$ in the adjoint
representation.  One may worry that after decoupling, one may lose
solutions at $k=0$, but a closer examination shows that the only
solution for $k=0$ is the trivial one where all fields are zero.  We
will see this after we derive the master equations.

In order to consider the construction of an $N{=}2$ SCA with this
data, and according to the result quoted in the introduction, one need
only write the most general $G^\pm(z)$
$$
G^\pm(z) = A_\pm^{ij} (I_i \psi_j)(z) +
B_\pm^{ijk}(\psi_i\psi_j\psi_k)(z) + C_\pm^i \d\psi_i(z)~,\(gg)
$$
where $B_\pm^{ijk}$ is totally antisymmetric and $A_\pm^{ij}$ is
arbitrary, and then derive the equations on the $\fr1/3 n^3 + n^2 +
\fr8/3 n + 1$ ($n=\dim\gg$) free parameters: $A_\pm^{ij}$,
$B_\pm^{ijk}$, $C_\pm^i$, and $k$ that come induced from \(GGope) and
\(JGope).

Imposing \(GGope) gives rise to the following equations.  From the
third order pole we find\fnote{Needless to say, we have used the
$\hbox{\sl Mathematica}^{\rm TM}$ package {\it OPEdefs}\/ written by
Kris Thielemans \[Kris].  I would like to take this opportunity to
thank him for his patient tutoring, and Steffen Mallwitz for access to
the computer.  Without their help, this would have taken much longer
and not been as enjoyable.}
$$
\eqnalign{
0={}& k\left(-2C_\pm^iC_{\pm\,i} + k A_\pm^{ij} A_{\pm\,ij} -
4 \f[ij,k] A\pm^{ij}C_{\pm\,k} - 6 k^2 B_\pm^{ijk}B_{\pm\,ijk}\right.\cr
&\left.- 6 k A_{\pm\,ij}B_\pm^{jkl}\f[kl,i] + (\f[ij,m]f_{mlk} -
\f[jl,m]f_{mik}) A_\pm^{ij}A_\pm^{kl}\right)~.\(poleiii)\cr}
$$
The equations from the double pole are much simpler:
$$
0 = A_\pm^{ij}A_\pm^{kl} \f[il,m]\f[jk,n] - \exc[m,n] ~.\(poleii)
$$
Finally let us consider the equations coming from the simple pole:
$$
\eqnalign{
0={}& k A_\pm^{ij} A_\pm^{kl} g_{jl} + \exc[i,j]\(poleia)\cr
0={}& k \left( A_\pm^{ij} A_\pm^{kl} f_{jkl} + \fr1/2 A_\pm^{mj}
A_\pm^{kl} g_{jl} \f[mk,i]+ k A_\pm^{ij} C_{\pm\,j}\right)\(poleib)\cr
0={}& 2 A_\pm^{mk} B_\pm^{ijn}\f[mn,l] + 3 k B_\pm^{ijm}B_\pm^{kln}
g_{mn} + \hbox{signed perms in }(i,j,k,l)\quad\(poleic)\cr
0={}& A_\pm^{ki}C_\pm^l\f[kl,j] + 6k A_\pm^{kl} B_\pm^{imn}\f[kn,j]
g_{lm} + 3k B_\pm^{ijk}C_\pm^l g_{kl}+ 3k A_\pm^{kl}
B_\pm^{ijm}f_{klm}\cr
&+ 9k^2 B_\pm^{ikl} B_\pm^{jmn} g_{kn} g_{lm} + 3k A_\pm^{kj}
B_\pm^{ilm}f_{klm} + \exc[i,j]\(poleid)\cr
0={}&6 k A_\pm^{il} B_\pm^{jkm}g_{lm} + A_\pm^{lj}A_\pm^{mk}\f[lm,i] -
2 A_\pm^{il}A_\pm^{mj}\f[lm,k] + \exc[j,k]\(poleie)\cr
0={}&\fr1/2 A_\pm^{kl}A_\pm^{mn}\f[kn,j]\f[lm,i] -
A_\pm^{kl}A_\pm^{mj}\f[km,n]\f[ln,i] + \fr{k}/2
A_\pm^{ki}A_\pm^{lj}g_{kl} + \exc[i,j]~.\(poleif)\cr}
$$
{}From \(GpGmope) we can read off the central charge
$$
\eqnalign{
c ={}& \fr{3k}/2 \left(- C_+^iC_{-\,i} + \fr{k}/2 A_+^{ij}A_{-\,ij}
- A_+^{ij}C_-^k f_{ijk} - 3k^2 B_+^{ijk}B_{-\,ijk} \right.\cr
& \left. + \fr{3k}/2 A_+^{ij}B_-^{klm} f_{ikl}g_{jm} - \fr1/2
A_-^{ij}A_+^{kl}\f[il,m]f_{jkm} - A_-^{ij}A_+^{kl}\f[ik,m] f_{jlm} +
\exc[{+},{-}] \right)~,\cr
&\(centralcharge)\cr}
$$
and the expression for $J(z)$.  We choose to write $J(z) = E^i I_i(z) +
F^{ij}(\psi_i\psi_j)(z)$, where $E^i$ and $F^{ij}$ can be found from
the double pole in \(GpGmope):
$$
\eqnalign{
E^i ={}& \fr{k}/2 \left( A_+^{ij} C_{-\,j} + A_-^{jk} A_+^{il} f_{jkl}
- \fr1/2 A_-^{jk}A_+^{lm} g_{km} \f[jl,i] - \exc[{+},{-}]\right)\(defe)\cr
F^{ij} ={}& \fr1/2 \left( A_+^{ki} C_-^l\f[kl,j] + \fr1/2
A_-^{kl}A_+^{mn} \f[kn,i]\f[lm,j] + A_-^{kl} A_+^{mi}
\f[km,n]\f[ln,j]+ \fr{k}/2 A_-^{kj}A_+^{li}g_{kl}\right.\cr
&\left. + 3k B_+^{ijk} C_{-\,k} + 3k A_-^{kl}B_+^{ijm} f_{klm} + 9k^2
B_-^{jkl}B_+^{imn} g_{kn}g_{lm} + 3k A_+^{ki} B_-^{jlm}f_{klm}\right.\cr
&\left. + 6k^2A_-^{kl} B_+^{imn}\f[kn,j] g_{lm} -
\exc[{+},{-}]\right)~.\(deff)\cr}
$$
Finally from \(JGope) one can obtain the final set of equations, which
we leave in terms of $E^i$ and $F^{ij}$.  The first set of equations
comes from the double pole
$$
\eqnalign{
0={}& 2k C_{\pm\,k}F^{ik} + 6k^2 B_\pm^{ijk}F_{kj} - 2k
A_\pm^{jk}F^{lm}\f[jl,i]g_{km} + 2k A_\pm^{jk}F^{il}f_{jkl}\cr
& + k A_\pm^{ji}F^{kl} f_{jkl}- C_\pm^{j}E^k \f[jk,i] + k
A_\pm^{ji}E_j + A_\pm^{jk}E_l \f[jl,m]\f[km,i] + 3k B_\pm^{ijk}E^l
f_{jkl}~,\cr
&\(poleiijg)\cr}
$$
whereas from the single pole come the following equations
$$
\eqnalign{
A_\pm^{ij} ={}& \pm 2k A_\pm^{ik} F^{jl} g_{kl} \mp A_\pm^{kj}E^l
\f[kl,i] \mp A_\pm^{ik}E^l\f[kl,j]\(poleijga)\cr
B_\pm^{ijk} ={}& \pm \fr1/3 A_\pm^{lj}F^{im} \f[lm,k] \pm k
B_\pm^{jkl}F^{im}g_{lm} \cr
& \mp \fr1/2 B_\pm^{ijl}E^m \f[lm,k] + \hbox{signed perms in
}(i,j,k)\quad\(poleijgb)\cr
C_\pm^{i} ={}& \pm 2k C_{\pm\,k}F^{ik} \mp 2k
A_\pm^{jk}F^{lm}\f[jl,i]g_{km} \pm 2k A_\pm^{jk}F^{il}f_{jkl} -
C_\pm^{j}E^k\f[jk,i]~.\cr
&\(poleijgc)\cr}
$$

The form of this last set of equations suggests a deformation-type
approach, in which by starting from a nontrivial solution of the
master equations we can find other solutions depending formally on
some parameter and agreeing with the original solution as the
parameter tends to zero.  This will undoubtedly miss many interesting
``nonperturbative'' solutions which only exist for a discrete set of
values of the parameter, but it has proven in the past to be a useful
method.  A natural deformation parameter is the level $k$.  However we
have to determine around which point to deform.  The only natural
points are $k=0$ and $k^{-1}=0$.  It is easy to see that for $k=0$,
the only solution is the trivial one.  In fact, from \(defe), we see
that $E^i=0$ at $k=0$.  Plugging this into \(poleijga) and \(poleijgc)
we see that $A_\pm^{ij}=0$ and $C_\pm^{i}=0$.  Finally into
\(poleijgb), we find that $B_\pm^{ijk}=0$, leading to a trivial
solution. Therefore the natural point from which to expand is
$k^{-1}=0$.  The high-level analysis of the Virasoro master equation
has been studied in \[highlevel].   We will, however, not deal any
further with the perturbative approach in this paper.

Having ruled out anything interesting at $k=0$ we are free to decouple
the fermions.  We derive the master equations for this case in the
next section.

\section{The Master Equations with Decoupled Fermions}

We assume that $k\neq 0$ and we shift the currents as in \(shiftedj).
The shifted currents $J_i(z)$ satisfy the following algebra
$$
\eqnalign{
J_i(z) J_j(w) &= \ope[2][k' g_{ij}] + \ope[1][{f_{ij}}^k J_k(w)] +
\reg~, \(affineKMdec)\cr
\noalign{\hbox{where $k' = k - \fr1/2 c_{\gg}$, and}}
J_i(z) \psi_j(w) &= \reg~, \(Jpsidec)\cr}
$$
while the fermions still satisfy \(fermions).  It will prove useful in
what follows to rescale the fermions so that they obey the following
OPE:
$$
\psi_i(z) \psi_j(w) = \ope[1][g_{ij}] + \reg~. \(fermionsren)
$$

The most general expression for $G^\pm(z)$ is given by
$$
G^\pm(z) = A_\pm^{ij} (J_i \psi_j)(z) +
B_\pm^{ijk}(\psi_i\psi_j\psi_k)(z) + C_\pm^i \d\psi_i(z)~,\(ggnew)
$$
where the currents have been shifted and the fermions rescaled.

Imposing \(GGope) we find that the double pole equation is trivially
satisfied, whereas from the third and first order poles we obtain:
$$
\eqnalign{
0={}& k' A_\pm^{ij}A_{\pm\,ij} - 2 C_\pm^{i}C_{\pm\,i} - 6 B_\pm^{ijk}
B_{\pm\,ijk}\(poleiiidf)\cr
0={}& A_\pm^{ik}A_\pm^{jl}g_{kl}\(poleiadf)\cr
0={}& A_\pm^{ij} C_{\pm\,j}\(poleibdf)\cr
0={}& B_\pm^{ijk}C_{\pm\,k}\(poleicdf)\cr
0={}& A_\pm^{lj}A_\pm^{mk}\f[lm,i] + 6
A_\pm^{il}B_\pm^{jkm}g_{lm}\(poleiddf)\cr
0={}& B_\pm^{ijm}B_\pm^{kln} g_{mn} + \hbox{signed perms
in~}(i,j,k,l)\quad\(poleiedf)\cr
0={}& k' A_\pm^{ki}A_\pm^{lj}g_{lk} + 18
B_\pm^{ikl}B_\pm^{jmn}g_{kn}g_{lm}~,\(poleifdf)\cr}
$$
which are to be compared to \(poleiii) and \(poleia)--\(poleif).

Computing \(GpGmope) we can read off the central charge
$$
c = \fr{3k'}/2 A_-^{ij}A_{+\,ij} - 3 C_-^i C_{+\,i} - 9 B_-^{ijk}
B_{+\,ijk}~,\(centralchargedf)
$$
and the expression for $J(z)$.  We again choose to write $J(z) = E^i
J_i(z) + F^{ij}(\psi_i\psi_j)(z)$, where $E^i$ and $F^{ij}$ can be
found from the double pole in \(GpGmope):
$$
\eqnalign{
E^i ={}& \fr1/2 A_+^{ij} C_{-\,j} - \fr1/4 A_-^{jk} A_+^{lm} \f[jl,i]
g_{km} - \exc[{+},{-}] \(defedf)\cr
F^{ij} ={}& \fr{k'}/4 A_-^{kj}A_+^{li} g_{kl} + \fr3/2
B_+^{ijk}C_{-\,k} - \fr9/2 B_-^{jkl} B_+^{imn}g_{km}g_{ln}-
\exc[{+},{-}]~,\(deffdf)\cr}
$$
which are to be compared with \(defe) and \(deff).

Finally, we use \(JGope) to obtain the final set of master equations.
{}From the double pole we obtain
$$
0 = 2 F^{ij}C_{\pm\,j} - 6 B_\pm^{ijk} F_{jk} + k' A_\pm^{ji}E_j~,
\(poleiijgdf)
$$
whereas from the first order pole we find
$$
\eqnalign{
A_\pm^{ij} ={}& \pm \left( 2 A_\pm^{ik} F^{jl} g_{kl} - A_\pm^{kj}E^l
\f[kl,i]\right) \(poleijgadf)\cr
B_\pm^{ijk} ={}& \pm \left(B_\pm^{jkl}F^{im}g_{lm} + \hbox{signed
perms in }(i,j,k)\right) \quad\(poleijgbdf)\cr
C_\pm^{i} ={}& \pm 2 C_{\pm\,j}F^{ij}\(poleijgcdf)\cr}
$$
which replace \(poleiijg)--\(poleijgc).

A few observations can already be made.  Contracting \(poleijgcdf)
with $C_{\pm\,i}$ and using the antisymmetry of $F^{ij}$ we find that
$C_{\pm\,i}C_\pm^i = 0$.  Similarly contracting \(poleiadf) with
$g_{ij}$ we get that $A_\pm^{ij} A_{\pm\,ij}=0$.  Therefore, the sole
content of \(poleiiidf) is simply $B_\pm^{ijk} B_{\pm\, ijk} = 0$,
which at the same time follows from \(poleifdf) by contracting with
$g_{ij}$ and using the previous observation.  Consequently,
\(poleiiidf) is rendered redundant.  In other words, only the
equations from the first pole are relevant.

If we specialize to $C_\pm^i= 0$, we can compare our equations with
those found in the appendix D of \[GHKO].  In fact, in \[GHKO] the
authors do not consider $N{=}1$ affine Lie algebras, but rather an
affine Lie algebra and free fermions in the fundamental representation
of (a real form of) $SO(n,\comps)$.  Despite this small difference,
the equations for one case can be read from the equations in the
other; and the comparison is possible.  A quick glance at the two sets
of equations allows us to identify the equations (D.6a)-(D.6f) in
\[GHKO] with our equations \(poleiadf), \(poleifdf), \(poleiddf),
\(poleiedf), \(poleijgadf), and \(poleijgbdf) respectively.  Equation
(D.5) in \[GHKO], which corresponds to the $N{=}1$ master equation,
finds no analogue here, since as shown in \[jmf] and \[getzler], it is
redundant.

These master equations seem to have sufficient structure to allow for
a general analysis and perhaps their solution.  Work on this is in
progress and we hope to report on this in the near future.  In the
next section, however, we look for solutions in the simple case of
$\gg=sl(2)$.  This will suggest an Ansatz for general $\gg$ that we
analyze in section 5.

\section{$\gg=sl(2)$, a Geometrical Approach}

We now study the solutions that can be constructed from $\gg = sl(2)$.
Without loss of generality we can rescale our generators such that
$g_{ij} = \delta_{ij}$ and such that $f_{ijk}=\epsilon_{ijk}$.  This
allows us to identify $sl(2)$ with $\comps^3$ in such a way that the
Lie bracket goes over to the cross product and the invariant metric
goes over to the dot product.  This mild geometrization will prove
very useful in solving the master equations for $sl(2)$.

Notice that $B_\pm^{ijk} = b_\pm \epsilon^{ijk}$, and contracting
\(poleijgcdf) with $\epsilon_{ijk}$ one finds that
$$
b_\pm = \pm 2 F^{ij}\delta_{ij} = 0~,\(betas)
$$
whence $B_\pm^{ijk} = 0$.

To analyze the rest of the equations, let us define a family of
vectors $\vec{V}_\pm^i\in\comps^3$, by $(\vec{V}_\pm^i)^j =
A_\pm^{ji}$.  We can then rewrite the equations involving $A_\pm^{ij}$
as geometric equations for these vectors.  For instance, in this
language \(poleiddf) becomes
$$0=\vec{V}_\pm^i \times \vec{V}_\pm^j\()$$
which immediately says that the vectors $\vec{V}_+^i$ and
$\vec{V}_-^i$ are separately collinear; in other words, there are
vectors $\vec{V}_\pm$ and complex numbers $\alpha_\pm$, $\beta_\pm$
and $\gamma_\pm$ such that
$$
\vec{V}_\pm^1 = \alpha_\pm \vec{V}_\pm~,\quad
  \vec{V}_\pm^2 = \beta_\pm \vec{V}_\pm~,\quad\hbox{and}\quad
\vec{V}_\pm^3 = \gamma_\pm\vec{V}_\pm~.\(coll)
$$
If we introduce two new vectors $\vec{W}_\pm \in\comps^3$, by
$\vec{W}_\pm = (\alpha_\pm,\beta_\pm,\gamma_\pm)^t$, then
it follows that $A_\pm^{ij} = V_\pm^i W_\pm^j$; in other words,
$A_\pm$ are decomposable.

In terms of these new variables, \(poleiadf) becomes
$$
0 = (\vec{W}_\pm\cdot\vec{W}_\pm)
\vec{V}_\pm\tensor\vec{V}_\pm~,\(vecone)
$$
which is easily seen to imply \(poleiiidf).  Since we don't want a
trivial solution, \(vecone) simply says that $\vec{W}_\pm$ are null:
$$
\vec{W}_\pm\cdot\vec{W}_\pm = 0~.\(zeronormw)
$$
This does not imply that they are zero, however, since they are
vectors in $\comps^3$ and the inner product is not hermitian, as it
does not involve complex conjugation.  Similarly, \(poleifdf)
also says that $\vec{V}_\pm$ are null away from $k'=0$
$$\vec{V}_\pm\cdot\vec{V}_\pm = 0~.\(zeronormv)$$
{}From now on, we assume that $k'\neq 0$.  We shall treat the case
$k'=0$ separately.  We will see, however, that nothing special happens
there.

If we now use \(poleiijgdf) and \(poleijgcdf) we find
$$
\vec{C}_\pm = \mp k' (\vec{E}\cdot\vec{V}_\pm) \vec{W}_\pm~.\(citoW)
$$
This equation, together with the fact that $\vec{W}_\pm$ are null
\(zeronormw) immediately implies \(poleibdf).

Now, on the one hand, we can rewrite \(defedf) as
$$
\vec{E} = \fr1/2 (\vec{C}_-\cdot \vec{W}_+) \vec{V}_+ - \fr1/2
(\vec{C}_+\cdot \vec{W}_-) \vec{V}_+ + \fr1/2 (\vec{W}_+\cdot
\vec{W}_-) \vec{V}_+\times \vec{V}_-~,\()
$$
which dotted with $\vec{V}_\pm$ becomes
$$
\vec{E}\cdot \vec{V}_\pm = \mp \fr1/2 (\vec{C}_\pm\cdot\vec{W}_\mp)
(\vec{V}_+\cdot\vec{V}_-)~.\(EdotV)
$$
On the other hand, \(deffdf) can be rewritten as
$$
F^{ij}= \fr{k'}/4 (\vec{V}_+ \cdot \vec{V}_-) \left( W_+^i W_-^j -
W_+^j W_-^i \right)~.\(geomf)
$$
In this guise, it is easy to see that
\(poleijgcdf) is equivalent to \(citoW) after using \(EdotV).

The last equation to be transcribed into this geometric
language---\(poleijgadf)---becomes after some manipulation
$$
\left( 1 - \fr{k'}/2 (\vec{V}_+ \cdot \vec{V}_-) (\vec{W}_+ \cdot
\vec{W}_-) \right) \vec{V}_\pm = \pm
\vec{E}\times\vec{V}_\pm~.\(lasteq)
$$
The RHS of this equation can be read off of
$$
\vec{E}\times \vec{V}_\pm = \fr1/2 (\vec{C}_\pm \cdot \vec{W}_\mp)
\vec{V}_+ \times \vec{V}_- \mp \fr1/2  (\vec{V}_+\cdot\vec{V}_-)
(\vec{W}_+\cdot\vec{W}_-)  \vec{V}_\pm ~.\(eXv)
$$
Plugging \(eXv) into \(lasteq) we find
$$
\left( 1 - \fr{k'-1}/2 (\vec{V}_+ \cdot \vec{V}_-) (\vec{W}_+ \cdot
\vec{W}_-) \right) \vec{V}_\pm = \pm \fr1/2 (\vec{C}_\pm \cdot
\vec{W}_\mp) \vec{V}_+ \times \vec{V}_-~.\(finaleq)
$$

We can now argue that $\vec{C}_\pm$ have to be zero.  Indeed, dot
\(finaleq) with $\vec{V}_\mp$.  On the RHS we get zero, but on the LHS
we find
$$
\left( 1 - \fr{k'-1}/2 (\vec{V}_+ \cdot \vec{V}_-) (\vec{W}_+ \cdot
\vec{W}_-) \right) (\vec{V}_+\cdot\vec{V}_-) = 0~.\()
$$
Now, this can have either of two solutions, either $ (\vec{V}_+ \cdot
\vec{V}_-) = 0$ or not.  We argue that it cannot be zero.  If it were,
then $\vec{C}_\pm=0$ because of \(citoW) and \(EdotV).  And feeding
this back into \(finaleq) we see that $\vec{V}_\pm=0$, yielding the
trivial solution.  Therefore, we must have that
$$
\fr{k'-1}/2 (\vec{V}_+ \cdot \vec{V}_-) (\vec{W}_+ \cdot
\vec{W}_-) = 1 ~.\(quadconst)
$$
If this is the case then the RHS of \(finaleq) is also zero, whence
crossing with $\vec{V}_+$, yields that $\vec{C}_\pm=0$ after using
\(EdotV) and \(citoW).

Therefore $\vec{C}_\pm=0$ and we are left with \(quadconst) as our
only equation.   Notice that for $k'=1$ there are no solutions except
the trivial one.  This restriction is not new.  In fact, in the
normalization of the algebra that we have used, the eigenvalue of the
quadratic casimir $\sum_i \ad X_i \ad X_i$ on the adjoint
representation is $-2$.  Thus, $k' = k + 1$, and $k'=1$ corresponds to
the case $k=0$ which we had ruled out previously.

In summary, the initial system has been reduced to the following:
Given four null vectors $\vec{W}_\pm$ and $\vec{V}_\pm$ satisfying
\(quadconst), we have an $N{=}2$ SCA.  A priori this would indicate
that there is a continuum of solutions, even after we pass to the
moduli space; that is, after we account for the freedom to change
basis in the algebra.
% To get an idea of the size of the solution space, fix the level
% $k'\neq 1$ and notice that four null vectors in $\comps^3$ define an
% 8-dimensional subvariety of $\comps^{12}$ whereas the constraint
% \(quadconst) restricts us to a 7-dimensional subvariety.  Of course,
% from the construction it follows that $\vec{W}_\pm$ were not
% general, but rather were such that they had a coordinate fixed.
% This corresponds to the fact that it is the tensor product that goes
% into the solution via the matrices $A_\pm$. This takes away two
% dimensions, leaving us with a five-dimensional variety of solutions.
% We don't expect all these solutions to be different, though.  We
% must look at the space one obtains after identifying points in the
% parameter space that are related by  automorphisms of the algebra;
% that is, at the moduli space.  In other words, we are still allowing
% for the freedom of changing basis in the algebra.  The adjoint group
% of $sl(2)$ is $SO(3,\comps)$.  Hence we have to quotient still by
% its action on our parameters $A_\pm^{ij}$. Let $M_i^j$ denote an
% element of $SO(3,\comps)$.  Its action on the $A_\pm^{ij}$ is given
% by $$ A_\pm^{ij} \mapsto M_k^i M_l^j A_\pm^{kl}~.\() $$  And it is
% plain that the master equations are covariant under these
% transformations.  The analysis of the moduli space is outside the
% scope of this paper, but a naive count says that at regular points,
% it is two-dimensional.  \comment{We should analyze this space.}
However, computing the central charge of the $N{=}2$ SCA, we find
from \(centralchargedf) that
$$ \eqnalign{ c &= \fr{3k'}/2  (\vec{W}_+\cdot\vec{W}_-)
(\vec{V}_+\cdot\vec{V}_-) \()\cr
\noalign{\hbox{which upon using \(quadconst) becomes}}
c &= {3k' \over k'-1} = 3 + {3\over k}~.\(ccsolution)\cr}
$$
Remarkably it only depends on the level!  This suggests that there is
essentially only one construction.

Let us try to understand this construction.  The decomposition of
$A_\pm^{ij}$ suggests that we first define currents $V_\pm(z) \equiv
V_\pm^i J_i(z)$ and fermions $\psi_\pm(z) = W_\pm^i\psi_i(z)$.  The
currents obey the following algebra
$$
V_+(z) V_-(w) = \ope[2][k' \vec{V}_+\cdot \vec{V}_-] +
                \ope[1][V_0(w)] + \reg~,\()
$$
where $V_0(w)\equiv (\vec{V}_+ \times \vec{V}_-)^i J_i(z)$.  Clearly then,
the rescaled currents $\widetilde{V}_\pm$ and $\widetilde{V}_0$, where
$\widetilde{V}_0 = \lambda^2 V_0$ and $\widetilde{V}_\pm = \lambda
V_\pm$ and $\lambda^2 = -2/(\vec{V}_+\cdot \vec{V}_-)$, obey an affine
$sl(2)$ at level $k^*\equiv -2k'$.  On the other hand, the fermions
$\psi_\pm$ obey
$$
\psi_+(z) \psi_-(w) = \ope[1][\vec{W}_+\cdot \vec{W}_-] + \reg ~, \()
$$
which becomes the standard OPE of a fermionic $bc$-system after
rescaling them by $(\vec{W}_+\cdot \vec{W}_-)^{-1/2}$.  Notice that
both the rescaling of the fermions and of the currents can be made,
since \(quadconst) forbids either $\vec{V}_+\cdot \vec{V}_-$ or
$\vec{W}_+\cdot \vec{W}_-$ from being zero.  Notice also that in terms
of the level $k^* = -2k'$ of the affine $sl(2)$, \(ccsolution) agrees
(for positive integer values of $k^*$) with the formula for the
unitary minimal series.  This suggests that this construction may be
related to the parafermionic constructions of the $N{=}2$ minimal
models.  We will see in Section 6 that this is indeed the case.

Thus the construction now becomes clear: given an $sl(2)$ at level
$k^*\neq -2$ and a fermionic $bc$-system, we can construct an $N{=}2$
SCA, with central charge given by \(ccsolution).  Furthermore, as we
will show in the next section, this construction admits an embedding
in any $N{=}1$ affine Lie algebra and also a mild generalization.  But
before, and as advertised, we discuss the case $k'=0$.

Equation \(poleifdf) does not imply now that $\vec{V}_\pm$ are null.
However, since $F^{ij} = 0$ now, then \(poleijgcdf) says that
$\vec{C}_\pm = 0$.  This means that $\vec{E} = \fr1/2
(\vec{W}_+\cdot\vec{W}_-) \vec{V}_+ \times \vec{V}_-$, and thus
equation \(poleijgadf) becomes, after discarding the trivial solution,
$$
\vec{V}_\pm = \pm \vec{E} \times \vec{V}_\pm~,\()
$$
which implies that $\vec{V}_\pm$ are indeed null.  Therefore, the
solution found before extends to $k'=0$ and is moreover the only
solution there.

\section{A Construction for general $\gg$}

We work with decoupled fermions, so the notation is as in Section 3.
We will take as our Ansatz
$$
G^\pm(z) =  A_\pm^{ij} (J_i\psi_j)(z)~,\(ansatz)
$$
where $A_\pm$ is decomposable.  That is, we consider elements $V_\pm =
\sum_i V_\pm^i X_i$ and $W_\pm=\sum_i W_\pm^i X_i$ in $\gg$ and define
$A_\pm^{ij} = V_\pm^i W_\pm^j$.

Equations \(poleiadf) says that $W_\pm$ are null; whereas, away
from $k'=0$, \(poleifdf) says that $V_\pm$ are null too.  Similar
considerations as in the previous section allow us to conclude that
nothing special happens for $k'=0$ and that $V_\pm$ are null there
too.  We shall omit the details this time.

If we define $E\in\gg$ by $E=\sum_i E^iX_i$ then \(defedf) can be
rewritten as
$$
E = \fr1/2 \pair<W_+,W_-> \comm[V_+,V_-]~,\(gene)
$$
whereas defining $F\in\bigwedge^2 \gg$ by $F = \fr1/2 \sum_{i,j}F^{ij}
X_i\wedge X_j$ turns \(deffdf) into
$$
F = \fr{k'}/4 \pair<V_+,V_-> W_+ \wedge W_-~.\(genf)
$$

Now, using \(genf) we can rewrite \(poleiijgdf) as $ k' W_\pm
\pair<E,V_\pm> = 0$.  Since $k'\neq 0$ and we are discarding the
trivial solution, this says that $\pair<E,V_\pm> = 0$.  Using \(gene)
this becomes
$$
\pair<E,V_\pm> = \fr1/2 \pair<W_+,V_->
\pair<{\comm[V_+,V_-]},V_\pm>~,\()
$$
which is already zero by invariance of $\pair<,>$.

Finally \(poleijgadf) becomes---discarding the trivial solution---
$$
\left( 1 - \fr{k'}/2 \pair<V_+,V_-> \pair<W_+,W_-> \right) V_\pm = \pm
\comm[E,V_\pm]~,\()
$$
which using \(gene) turns into
$$
\left( 1 - \fr{k'}/2 \pair<V_+,V_-> \pair<W_+,W_-> \right) V_\pm = \pm
\fr1/2 \pair<W_+,W_-> \comm[{\comm[V_+,V_-]},V_\pm]~.\(sltwoid)
$$
This equation may seem formidable, but it is actually easy to
recognize.  Let us define $V_0 \equiv \comm[V_+,V_-]$.  Then
\(sltwoid) becomes
$$
\comm[V_0,V_\pm] = \pm 2 {\left( 1 - \fr{k'}/2 \pair<V_+,V_->
\pair<W_+,W_-> \right) \over \pair<W_+,W_->} V_\pm~.\(almostsltwo)
$$
Now, we can always choose $W_\pm$ such that the RHS is not zero and
also finite.  In fact, we can always divide by $\pair<W_+,W_->$, since
if it were zero, then \(sltwoid) would say that $V\pm=0$ rendering the
solution trivial.  Similarly, for any fixed value of the level, one
can rescale $W_\pm$ in such a way that the RHS of \(almostsltwo) is
not zero.  Therefore for those generic\fnote{It would remain to study
those $W_\pm$ for which $\pair<W_+,W_-> = 2/ k' \pair<V_+,V_->$.
These correspond to embeddings in $\gg$ of the Lie algebra defined by
$\comm[V_+,V_-] = V_0$ and $\comm[V_0,V_\pm]=0$.  For $\gg=sl(2)$ such
embeddings clearly do not exist; but they do exist for other simple
$\gg$.  It may be interesting to classify them.} $W_\pm$,
\(almostsltwo) simply says that suitably rescaling $V_\pm$ and $V_0$
they span an $sl(2)$ subalgebra of $\gg$.  Explicitly, if we let
$\widetilde{V}_\pm = \lambda V_\pm$ and $\widetilde{V}_0 = \lambda^2
V_0$, where
$$
\lambda^2 = { 2 \pair<W_+,W_-> \over 2 - k' \pair<W_+,W_->
\pair<V_+,V_->}~,\()
$$
then $\comm[\widetilde{V}_0,\widetilde{V}_\pm]= \pm 2
\widetilde{V}_\pm$ and $\comm[\widetilde{V}_+,\widetilde{V}_-] =
\widetilde{V}_0$.

Conversely, given any $sl(2)\subset\gg$, can one find null vectors
$W_\pm$ such that we obtain an $N{=}2$ SCA?  It follows from the
invariance of $\pair<,>$ that if $\widetilde{V}_\pm$ and
$\widetilde{V}_0 \equiv \comm[\widetilde{V}_+,\widetilde{V}_-]$
satisfy an $sl(2)$ then $V_\pm$ are necessarily null.  Thus the
question reduces to whether we can find null vectors $W_\pm$ such that
$$
1 - \fr{k'}/2 \pair<V_+,V_-> \pair<W_+,W_->  =  \pair<W_+,W_->~, \()
$$
where $V_\pm$ are suitable rescalings of $\widetilde{V}_\pm$ which,
together with a rescaling $V_0$ of $\widetilde{V}_0$, still obey an
$sl(2)$.  The answer to this is clearly affirmative, since we can
always rescale $V_\pm$ in such a way that $k'\pair<V_+,V_-> \neq -2$,
whatever the value of $k'$.

Moreover, notice that from \(centralchargedf), we can obtain the
central charge:
$$
c = {3k' \pair<V_+,V_-> \over 2 + k' \pair<V_+,V_->}~.\(ccgen)
$$
Notice that $k' \pair<V_+,V_->$ is the induced level $k^*$ of $sl(2)
\subset \gg$ -- the index of embedding need not be 1.  As before, we
recover the minimal unitary series for $k^*$ a positive integer.

Finally, let us remark that for $\gg \neq sl(2)$ the construction
affords a slight generalization, since one can have that $C_\pm \neq
0$.  It is easy to see that if we can find elements $C_\pm\in\gg$
obeying
$$
\pair<C_\pm,C_\pm> = \pair<C_\pm,W_\pm> =
\pair<C_\pm,W_\mp>=0~,\(cconds)
$$
then we obtain an $N{=}2$ SCA with central charge
$$
c = {3k' \pair<V_+,V_-> \over 2 + k' \pair<V_+,V_->} - 3\pair<C_+,C_->
{}~.\(ccgendef)
$$
Notice that it is impossible to find nonzero vectors
$C_\pm\in\comps^3$ obeying \(cconds), which is why this deformation of
the construction does not exist for $\gg=sl(2)$.

\section{Relation to Kazama--Suzuki Models and Parafermions}

In this section we will identify the unique construction for $sl(2)$,
with the Kazama--Suzuki coset construction associated to
$SO(3)/SO(2)$, with a (generalized) parafermionic construction and
also with a bosonic construction of the Kazama--Suzuki type.

\subsection{Kazama--Suzuki Coset Construction}

As we made explicit in the previous two sections, the construction of
the $N{=}2$ SCA derives from the following one.  Let us start with an
$sl(2)$ affine Lie algebra at level $k^*$ with currents $J_\pm(z)$,
$J_0(z)$ obeying the OPEs
$$
\eqnalign{
J_0(z)J_\pm(w) &= \ope[1][\pm 2J_\pm(w)] + \reg~,\cr
J_+(z)J_-(w) &= \ope[2][k^*] + \ope[1][J_0(w)] + \reg~, \(sltwoKM)\cr
J_0(z)J_0(w) &= \ope[2][2k^*] + \reg~;\cr}
$$
and we then add a fermionic $bc$-system $\psi_\pm(z)$ with OPE
$$
\psi_+(z) \psi_-(w) = \ope[1][1] + \reg~. \()
$$
Then the following generators
$$
\eqnalign{
J ={}& {1\over k^*+2} \left[ J_0 + k^* (\psi_+\psi_-)\right]~,\()\cr
G^\pm ={}&  {1\over\sqrt{k^*+2}}  (J_\pm\psi_\pm)~,\(brst)\cr
\noalign{\hbox{and}}
T ={}&  {1\over k^*+2} \left[ (J_+J_-) + (J_0\psi_+\psi_-) -
\fr1/2 \d J_0 + \fr{k^*}/2 \left( (\d\psi_+\psi_-) -
(\psi_+\d\psi_-)\right) \right]~,
\cr
&\()\cr}
$$
satisfy an $N{=}2$ SCA with central charge given by
$$
c = {3k^* \over k^*+2}~.\()
$$
The form of $G^\pm(z)$ agrees (modulo conventions) with the
Cartan--Weyl basis description of the Kazama--Suzuki coset associated
to the 2-sphere as the hermitian symmetric space $SO(3)/SO(2)$---see,
for example, equation (4.4) in \[KaSu].  Since the $G^\pm(z)$
determine the whole SCA, we conclude that the two constructions are
identical.

\subsection{Parafermionic Construction}

One can gain some more insight into this construction by paying close
attention to $T(z)$.  It is easy to see that it is made out of two
commuting pieces: $T(z) = T_1(z) + T_2(z)$. The first term is
nothing but the coset construction $sl(2)/gl(1)$:
$$
T_1(z) = T_{sl(2)}(z) - T_{gl(1)}(z)~,\(cosetT)
$$
with
$$
T_{sl(2)} = {1\over 2(k^* + 2)}\left[ (J_+J_-) + (J_-J_+) + \fr1/2
(J_0J_0) \right]\(Tsuga)
$$
the Sugawara tensor, and
$$
T_{gl(1)} = {1\over 4k^*} (J_0J_0)\(glone)
$$
the $gl(1)$ piece.  And the second term is another $gl(1)$ coming from
$J(z)$:
$$
T_2(z) = {k^*+2 \over 2k^*} (JJ)(z)~.\(gltoo)
$$
This split suggests that the realization is nothing but a generalized
(that is, for not necessarily integer level) parafermionic
construction \[Gepner] \[Ahn], where the $N{=}2$ theory is equivalent
to the tensor product of a generalized parafermionic theory and a free
boson. Indeed, as we now show, bosonizing the $\psi_\pm$ system, one
can easily see that the two theories are the same.

Let us first rewrite the $sl(2)$ currents in terms of generalized
parafermions $\chi^{\phantom{\dagger}}_i$, $\chi^\dagger_i$ and a free
boson $\phi$ normalized to $\phi(z) \phi(w) = -\log (z-w) + \cdots$.
The affine currents are then written as
$$
\eqnalign{
J_0 &= i\sqrt{2k^*} \d\phi~,\cr
J_+ &= \sqrt{k^*} \chi^{\phantom{\dagger}}_1 \exp (i\sqrt{\fr2/{k^*}}
\phi)~,\(paraff)\cr
J_- &= \sqrt{k^*} \chi_1^\dagger \exp (-i\sqrt{\fr2/{k^*}} \phi)~,\cr
}
$$
with the parafermion fields obeying
$$
\chi^{\phantom{\dagger}}_1(z)\chi_1^\dagger(w) = {1\over
(z-w)^{2\Delta_1}} \left[ 1 + {2\Delta_1\over c_\chi} (z-w)^2 T_1(w) +
\cdots\right]\(parafope)
$$
with $T_1$ the coset energy-momentum tensor given by \(cosetT),
$\Delta_1 = \fr{k^*-1}/{k^*}$, and $c_\chi = \fr{2k^*-2}/{k^*+2}$ the
central charge of the coset theory.  Other parafermionic OPEs involve
further parafermionic pairs $\chi^{\phantom{\dagger}}_i$ and
$\chi^\dagger_i$ of dimension $\Delta_i = \fr{i(k^*-i)}/{k^*}$.  For
$k^*$ a positive integer we can truncate the spectrum getting rid of
all fields of negative dimension by the constraint $\chi^\dagger_i =
\chi^{\phantom{\dagger}}_{k^*-i}$; but for arbitrary $k^*$ these
currents are all there.

We now introduce a second scalar field $\xi$ with the same
normalization as $\phi$, and we use it to bosonize the $bc$-system
$\psi_\pm$ by
$$
\psi_\pm = \exp ( \pm i \xi )\qquad\hbox{and}\qquad (\psi_+\psi_-) = i
\d\xi~. \(bosonize)
$$
In terms of the two bosons and the parafermions, the $N{=}2$ currents
can now be written as
$$
\eqnalign{
J ={}& i\sqrt{\fr{k^*}/{k^*+2}} \d\pi~,\(jpf)\cr
G^+ ={}&  \sqrt{\fr{k^*}/{k^*+2}} \chi^{\phantom{\dagger}}_1\, \exp (i
\sqrt{\fr{k^*+2}/{k^*}} \pi) ~,\(gppf)\cr
G^- ={}&  \sqrt{\fr{k^*}/{k^*+2}} \chi_1^\dagger\, \exp (-i
\sqrt{\fr{k^*+2}/{k^*}} \pi) ~,\(gmpf)\cr
\noalign{\hbox{and}}
T ={}&  T_1 - \fr1/2 \d\pi\d\pi ~,\(tpf)\cr}
$$
where $\pi \equiv \sqrt{\fr2/{k^*+2}} (\phi + \sqrt{\fr{k^*}/2}\xi)$.
Thus the expression of the $N{=}2$ currents agrees with the ones for
the generalized parafermionic constructions as written, for example,
in equation (4.3) of \[Ahn] -- up to a rescaling of $\pi$.

\subsection{Bosonic Construction}

Alternatively, notice that both $T_{gl(1)}$ and $T_2$ have central
charge equal to 1.  Hence they can be bosonized by a free boson
without background charge.  In fact, if we bosonize the $sl(2)$
currents as well we see that the whole construction takes the form of
a bosonic $N{=}2$ construction of the type analyzed by Kazama and
Suzuki in \[KSBosonic].  There are at least two ways to bosonize the
$sl(2)$ currents: we could first introduce the Wakimoto representation
in terms of a free boson and a bosonic $(\beta,\gamma)$-system; and
then bosonize the $(\beta,\gamma)$-system in terms of another two free
bosons.  Alternatively, we can simply use \(paraff) and bosonize the
parafermion fields in terms of two extra bosons.  The bosonization of
generalized parafermions was considered in \[Ahn] and we will roughly
follow their approach.

Let us introduce free bosons $\varphi_i(z)\varphi_j(w) = - \delta_{ij}
\log(z-w) + \cdots$.  In \[Ahn] and references therein, the
parafermion currents $\chi_1^{\phantom{\dagger}}$ and $\chi_1^\dagger$
are shown to admit the following bosonization:
$$
\eqnalign{
\chi^{\phantom{\dagger}}_1 &= {1\over \sqrt{2}} \left[ \d \varphi_2 -
i \sqrt{\fr{k^* + 2}/{k^*}} \d \varphi_1 \right] \exp
(\sqrt{\fr2/{k^*}} \varphi_2)\()\cr
\noalign{\hbox{and}}
\chi_1^\dagger &= -{1\over \sqrt{2}} \left[ \d \varphi_2 + i
\sqrt{\fr{k^* + 2}/{k^*}} \d \varphi_1 \right] \exp
(-\sqrt{\fr2/{k^*}} \varphi_2)~.\()\cr}
$$
Plugging these formulas into \(jpf)--\(tpf), and letting $\varphi_3
\equiv \pi$, we find that the $N{=}2$ currents are given by
$$
\eqnalign{
J ={}& i\sqrt{\fr{k^*}/{k^*+2}} \d\varphi_3~,\(jbos)\cr
G^\pm ={}& {-i\over \sqrt{2}} \left[ \sqrt{\fr{k^* + 2}/{k^*}} \d
\varphi_1 \pm i \d\varphi_2 \right] \exp \pm (\sqrt{\fr2/{k^*}}
\varphi_2 + i \sqrt{\fr{k^*+2}/{k^*}} \varphi_3)~,\(gpmbos)\cr
\noalign{\hbox{and}}
T ={}&  - \fr1/2 \d\vec{\varphi}\cdot\d\vec{\varphi} + i
\sqrt{\fr2/{k^*+2}} \d^2\varphi_1 ~.\(tbos)\cr}
$$

\section{Conclusions}

Let us recapitulate the highlights of this paper.  Based on the
observation that only a few of the OPEs in the $N{=}2$ SCA are
sufficient to determine the rest, we have written down the minimal set
of $N{=}2$ master equations starting from an $N{=}1$ affine Lie
algebra with data $(\gg,k)$.  We saw that for $k=0$ there was no
nontrivial solutions to the master equations, whence we could decouple
the fermions without loss of generality.  After decoupling, the
equations simplify tremendously.  The set of master equations is given
by equations \(poleiadf)--\(poleifdf) coming from the first order pole
in the OPE $G^\pm(z)G^\pm(w)$; and by equations
\(poleiijgdf)--\(poleijgcdf) coming from the OPE $J(z)G^\pm(w)$.

After a mild but helpful geometrization of the master equations, we
solved them in general for the simplest case of $\gg=sl(2)$.  The
solution was shown to be equivalent to a construction out of an affine
$sl(2)$ at a shifted level, and a fermionic $bc$-system and
corresponding to the Kazama--Suzuki coset construction on the
hermitian symmetric space $SO(3)/SO(2)$.  Bosonizing the $bc$-system
we could also identify the construction with a generalized
parafermionic construction.  We also remarked that after bosonizing
the affine currents as well, the construction could be understood also
as a bosonic construction of the ones analyzed by Kazama and Suzuki in
\[KSBosonic].

Moreover, we saw that these results extend to some extent to general
$\gg$ and we showed that all solutions that fit the Ansatz
\(ansatz)---or even the more general Ansatz where we allow for
derivatives of the fermions---are related to embeddings
$sl(2)\subset\gg$ or to degenerations thereof (see the footnote in
Section 5).

Embeddings $sl(2)\subset\gg$ are also the building blocks---via
the (generalized) Drinfel'd--Sokolov (DS) reduction---of
$\W$-algebras.  Thus it behooves one to investigate the relation that
could exist between DS reduction and $N{=}2$ SCAs.  Some $N{=}2$
$\W$-superalgebras can be obtained by a DS-type reduction of affine
Lie superalgebras, but we have here something else in mind.  From
equation \(brst) it follows that the charge associated to the current
$G^+(z)$ can be interpreted as the BRST operator associated to the
constraint $J^+ = 0$.  This is to be contrasted with the constraint
imposed in the DS reduction $J^+ = 1$.  These comments suggest the
following conjecture: that associated to the $\W$-algebra coming from
the Drinfel'd--Sokolov reduction with data $sl(2)\subset\gg$ there is
an $N{=}2$ $\W$-superalgebra whose spectrum (in terms of $N{=}2$
supermultiplets) agrees with the spectrum of $\gg$ as irreducible
representations of $sl(2)$.  For the special case of the principal
embeddings $sl(2) \subset A_n$, these $N{=}2$ $\W$-superalgebras do
exist (see, for example, \[FiRa]).  The proof of this conjecture,
should it hold true, is work in progress.

Finally, the uniqueness of the generalized construction for
$\gg=sl(2)$ suggests that the $N{=}2$ structures inside $N{=}1$ affine
Lie algebra are much more rigid than the Virasoro structures inside
affine Lie algebras, for already in the case of $sl(2)$ one has a rich
structure of generalized Virasoro constructions.  Should this rigidity
persist, the generalized $N{=}2$ constructions on $N{=}1$ affine Lie
algebras may allow for a classification.

\ack

It is a pleasure to thank Ezra Getzler, Marty Halpern, Noureddine
Mohammedi, and Niels Obers for their many comments and discussions;
and Ralph Blumenhagen and Michael Terhoeven for sharing some of their
knowledge of the $N{=}2$ literature.  I am also thankful to the
Department of Physics of Queen Mary and Westfield College for their
hospitality during the final stages of this work.

\refsout
\bye